\documentclass[conference]{IEEEtran}

\usepackage{cite}
\usepackage{amsmath,amssymb,amsfonts,amsthm}
\usepackage{graphicx}
\usepackage{textcomp}
\usepackage{xcolor}
\def\BibTeX{{\rm B\kern-.05em{\sc i\kern-.025em b}\kern-.08em
    T\kern-.1667em\lower.7ex\hbox{E}\kern-.125emX}}

\usepackage{eso-pic}
\usepackage{hyperref}
\usepackage{algorithm}
\usepackage[noend]{algpseudocode}
\usepackage{array, makecell} 
\usepackage{multirow}
\usepackage{epstopdf}
\usepackage{ulem}
\usepackage{caption}
\usepackage{footnote}
\usepackage{booktabs}
\usepackage{mathrsfs, mathtools}

\usepackage{authblk}

\theoremstyle{definition}

\newtheorem{theorem}{Theorem}[section]

\newtheorem{lemma}[theorem]{Lemma}

\newtheorem{definition}[theorem]{Definition}

\newtheorem{corollary}[theorem]{Corollary}

\usepackage{cancel}

\usepackage{pifont}

\def\bitcoin{%
  \leavevmode
  \vtop{\offinterlineskip 
    \setbox0=\hbox{B}%
    \setbox2=\hbox to\wd0{\hfil\hskip-.03em
    \vrule height .3ex width .15ex\hskip .08em
    \vrule height .3ex width .15ex\hfil}
    \vbox{\copy2\box0}\box2}}

\newcommand{\commit}{Hash}
\newcommand{\commitment}{\textit{com}}
\newcommand{\commitments}{\textit{Com}}
\newcommand{\commRandomness}{\textit{r}}
\newcommand{\commSerialNum}{\textit{S}}

\newcommand{\cryptoParam}{\kappa}

\newcommand{\height}{H}

\newcommand{\MIFP}{\epsilon}

\newcommand{\player}{\textit{P}}

\newcommand{\protocolName}{\text{PACCs}}

\newcommand{\revealTXDelay}{\Delta}

\begin{document}

\title{Private, Anonymous, Collateralizable Commitments vs. MEV (Extended Version)}

\author[1]{Conor McMenamin}
\author[2]{Vanesa Daza}
\author[3]{Xavier Salleras}
\affil[1,2,3]{Department of Information and Communication Technologies, Universitat Pompeu Fabra, Barcelona, Spain}
\affil[1]{Nethermind Research}
\affil[2]{CYBERCAT - Center for Cybersecurity Research of Catalonia}
\affil[3]{Dusk Network}

\maketitle

\begin{abstract}
    In this work, we introduce the private, anonymous, collateralizable commitments (PACCs) framework. PACCs allow any smart contract wallet holder to collateralize a claim, request, or commitment in general, in a private and anonymous manner. PACCs can prove arbitrarily much or little about the wallet generating the commitment, and/or the transaction which is being committed. 
    We demonstrate that PACCs can be applied to effectively eliminate maximal-extractable value (MEV) in DeFi where it currently occurs, shifting MEV instead to censorship. After describing our protocol with detail, we provide an implementation using the Ethereum blockchain, and whose benchmarks prove how PACCs are completely feasible.
\end{abstract}

\begin{IEEEkeywords}
Zero-Knowledge Proofs, MEV, Commitments, Decentralized Finance
\end{IEEEkeywords}

\section{Introduction}

Miner-/maximal-extractable value (MEV) losses on Ethereum are upwards of $\$500M$ \cite{QuantifyingExtractableValueGervais,FlashbotsExploreWebsite}, with actual losses across all blockchains likely in the billions. These losses are being incurred by protocol users. If genuine protocol users are losing money, this diverts funds away from protocols to extractors, entities solely interested in investing in extraction techniques, while standing as a clear deterrent to prospective users. This is a direct consequence of users announcing an intent to do something (adding an unencrypted transaction to the mempool) before that action is executed on-chain. In the two primary sources of MEV, decentralized exchange (DEX) and auctions (liquidation auctions, non-fungible token auctions, etc.), this intent typically reveals some combination of order price, order size, sender identity, trading history, and account balances. All of these pieces of information allow extractors to extract value from players in a blockchain. 

Although some protocols have emerged in literature to hide some of this information \cite{P2DEXBaum,BlockAuctionPeriodicAuctionsConstantinides,AequitasKelkar,PubliclyVerifiableSecrecyPreservingPeriodicAuctionsGalal,HelixAsayag,EagleBaum,FairTraDEXMcMenamin}, no satisfactory decentralized solution has been found. One emergent technology which effectively hides transaction information on public blockchains is that of zero-knowledge mixers as used in \cite{SemaphoreWhitehat,FairTraDEXMcMenamin}. These allow players to join an anonymity set and send/commit to send transactions by (only) revealing membership in a set. Unfortunately, existing solutions require players to join anonymity sets in advance of proving membership, and for many other players to join the set to adequately hide identity. If protocols have disjoint anonymity sets, which all of the listed protocols do, this requires a player to restrict tokens to one of these protocols in advance of using the tokens. Furthermore, on-boarding and off-boarding into these anonymity sets require costly on-chain operations (merkle-tree additions, proof-verification), serving as bottlenecks to prospective protocols based on these anonymity sets.

\subsection{Our Contribution}

In this paper, we address the limitations of single-use anonymity sets, and their potentially costly waiting times for collateral to be usable. We propose an improved variant of single-use collateralized commitments, as used in \cite{FairTraDEXMcMenamin,LibSubmarine},  with a dynamic variation where collateral can be locked instantaneously for its required purpose, with commitments possible for almost any transaction, while effectively keeping that transaction private (no information about the purpose of the transaction itself is leaked) and anonymous (there is no way to learn the identity of the user initiation such transaction). These crucial improvements have immediate consequences for MEV, minimizing MEV opportunities as they currently occur in DEXs, auctions, and beyond. 

We describe \textit{Private, Anonymous, Collateralizable Commitments} (PACCs), a commitment protocol based on smart contract wallets (SCWs) and Zero-Knowledge Proofs (ZKPs). PACCs can be used to convince a prospective block builder or relayer that the user generating the PACC has enough funds to pay required fees, that its wallet is committed to performing certain actions, and importantly, that the wallet loses some amount of collateral if this commitment is broken. Our protocol performs expensive computing operations off-chain (i.e. computing ZKPs), only requiring few additional mapping checks when compared to transactions being sent from basic externally owned accounts (EOA), as in Ethereum \cite{Ethereum}. Mappings are gas-efficient storage structures implemented on smart contract-enabled blockchains, and are used in our construction to enforce commitments. 

We outline the properties of PACCs (Section \ref{sec:properties}), and demonstrate how they can be applied to effectively eliminate MEV as it is known in DEXs, liquidations and auctions. PACCs shift MEV to censorship, which is still a concern. However, we believe censorship can be made arbitrarily expensive \cite{MultiBlockProposerProbabilities,proposercensorshipResnick} by forcing protocols to accept commitments for long enough (we introduce a parameter  $\revealTXDelay$ for this \textit{long enough} in the model assumption of Section \ref{sec;protocol}), and by making all transactions practically indistinguishable, something which PACCs achieves. 

Furthermore, we implement a PACCs proof-of-concept\footnote{https://github.com/xevisalle/paccs} allowing us to gather information about our solution (e.g. smart contract gas consumption, computing times), that proves its feasibility.

\subsection{Organization of the Paper}

Section \ref{sec:RW} analyzes previous work related to the information hiding in the context of MEV-protection. Section \ref{sec:ZK} introduces the cryptographic primitives needed to formally reason about $\protocolName$. Section \ref{sec;protocol} defines the  PACC protocol, and Section \ref{sec:im} explains the proof-of-concept we implemented, along with the benchmarks we performed. Some general properties of PACCs are outlined in Section \ref{sec:properties}. Section \ref{sec:sims} demonstrates the potential of PACCs in protecting against MEV, specifically with respect to decentralized finance. 

We conclude, with some further discussion on PACCs, in Section \ref{sec:conclusion}.

\section{Related Work}\label{sec:RW}

Related to the concept of hiding transaction information before it is committed to the blockchain are Eagle \cite{EagleBaum}, P2DEX \cite{P2DEXBaum}, and Penumbra \cite{Penumbra}. Each of these protocols use trusted committees to keep order information hidden. If the committee colludes, all trade information is revealed. As sophisticated committees can choose when and how to use this information, users can be convinced that MEV is not occurring. Although PACCs may be replicated using these committee-based protocols, PACCs are intended to be implemented using decentralized underlying protocols free from committee-based dependencies. User-run commit-reveal \cite{LibSubmarine,FairTraDEXMcMenamin}, or revealing using delay encryption \cite{DelayEncryptionBurdges,FairPoSChiang} offer potential solutions in this regard, both of which are compatible with PACCs. 

With respect to Eagle, P2DEX, and LibSubmarine \cite{LibSubmarine}, any tokens committed to protocols run by the committee are known. These inputs are indicators of imbalances in upcoming protocols, and can be used by all players in the system to extract value from these protocols, and as such, the users. Combating this requires many transactions in every time slot to sufficiently hide the imbalance signals produced by individual users. This brings risk for early senders in every time slot, which stands as a barrier for adoption. With PACCs, commitments do not require tokens to be sent before revealing, while still ensuring these tokens must be used in the protocol, as attested to by the commitment.

FairTraDEX \cite{FairTraDEXMcMenamin} proposes hiding trade information using non-interactive zero-knowledge proofs and anonymity sets such as those used in \cite{ZCash}. A player wanting to participate in a FairTraDEX auction is required to join an anonymity set specific to that auction some time before an auction begins, waiting until the user is sufficiently hidden within the anonymity set before submitting an order. Joining an anonymity set and restricting one's tokens to a single use far in advance of using the tokens all have costs for users, making the practicality of the MEV protection guarantees in \cite{FairTraDEXMcMenamin} limited. 

Flax \cite{FlaxDai} allows users to anonymize sends, resembling work on burn addresses \cite{LibSubmarine}. This is a definite improvement on the basic externally-owned account model where users typically send all transactions from one address. However, with regard to DEXs and auctions, price, size and direction are all revealed in Flax. As such, it provides minimal improvements with respect to MEV protection.

\section{Preliminaries}\label{sec:ZK}

\subsection{Zero-Knowledge Proofs}

Zero-Knowledge Proofs (ZKPs) are cryptographic primitives allowing a party (the prover) to prove to another party (the verifier) that a given information is true without leaking any secret information that is willed to remain private. In particular, we are interested in zk-SNARKs (Zero-Knowledge Succinct Non-Interactive Argument of Knowledge) \cite{PairingBasedNIZKsGroth}, a specific type of ZKP that allows provers to prove some information without further interaction beyond sending a single message. Plus, the proofs can be verified very efficiently, and this fact is crucial as our protocol needs a single party (the relayer) to be continuously verifying proofs sent by multiple users.

We do not specify which specific ZKP scheme to use, as the exact choice will depend on other efficiency factors, as well as resource limitations and/or the strength of the security assumptions taken by each scheme.

The set of operations that we want to prove in our protocol are described by an \textit{arithmetic circuit}. Those circuits also describe which values are public and which need to remain private. These are the algorithms describing a generic ZKP scheme:

\begin{itemize}
    \item  $(pk, vk) \leftarrow \textit{Setup}(circuit, 1^{\lambda})$: For a given arithmetic circuit \textit{circuit} describing what the ZKP has to prove, and a security parameter $\lambda$, outputs a proving key $pk$ and a verifying key $vk$.
    
    \item  $\pi \leftarrow$ \textit{Prove}$(pk, inputs)$: Given a proving key $pk$, a non-deterministic ZKP $\pi$ proving the circuit satisfiability for a given set of \textit{inputs} is returned. 
    
    \item  $0/1 \leftarrow$ \textit{Verify}$(vk, \pi)$: Given a proof $\pi$ and a verifying key $vk$, returns 1 if the proof is correct, and 0 otherwise.
\end{itemize}

\subsection{Membership Proofs}

This section outlines the tools\cite{ZKProofsSetMembershipBenarroch} used throughout this paper to prove membership of a commitment into a given set. We define a generic hash function \textit{Hash} acting as a commitment scheme, and a membership proof scheme \textit{SetMembership}. 
 
 \begin{itemize}
     \item $com \coloneqq \textit{Hash}(m)$: A deterministic, collision-resistant hash function taking as input a string $m \in \{ 0,1\}^{*}$, and outputting a string $\commitment \in \{0,1\}^{\Theta(\cryptoParam)}$.
     
     \item $\textit{merkle\_proof} \coloneqq \textit{SetMembership}(\commitment, \commitments)$: Given a Merkle tree of commitments $\commitments$, generates a membership proof \textit{merkle\_proof} that $\commitment \in \commitments$.
 \end{itemize}

The above stated hash function needs to be executed in-circuit, in order to prove knowledge of the preimage using a ZKP. As such, the choice of the specific scheme needs to be done considering it to be ZKP-friendly, meaning that can be proved efficiently.

\subsection{Smart Contract Wallets}

This section introduces smart contract wallets as they are used in this paper. Users using smart contract wallets instead of externally-owned accounts to increase account functionality has been proposed for use in many Ethereum Improvement Proposals, most notably in the now standardized ERC-4337 \cite{EIP4337}. It suffices to consider smart contract wallets as extensions of externally-owned accounts, on which we can apply additional constraints.  For ease of notation, we shorten smart contract wallet to wallet for the remainder of the paper.  

To reason about our framework, we assume some finite number of token denominations $n$, with the total quantity of tokens in the system denoted $T \in \mathbb{Z}^n_{+}$.  For simplicity, we assume all tokens in the system are contained in the set of wallets. For $W$ the set of all wallets, and some wallet $w \in W$, $w$ can be considered as a set of tokens. The function $bal(w) := [v_1,...,v_n]$ indicates that there are exactly $v_i$ token $i$'s in wallet $w$, $\forall i \in [1,...,n]$. As such, $\sum_{w\in W}bal(w)=T$. With this in hand, we introduce two distinct commitment mappings:
\begin{itemize}
    \item For every wallet $w \in W$, a \textit{secret commitment mapping} $C^w_\textit{scrt}: \{0,1\}^{\Theta(\cryptoParam)} \rightarrow v \subseteq w$. $C^w_\textit{scrt}(x)$ is such that if $x \neq y$, $C^w_\textit{scrt}(x) \cap C^w_\textit{scrt}(y) = \emptyset$. This $C^w_\textit{scrt}(x)$ is a mapping of secrets to mutually exclusive subsets of tokens in the wallet $w$. 
    \item A \textit{global transaction commitment mapping} $C_{TX}: \{0,1\}^{\Theta(\cryptoParam)} \rightarrow \{0,1\}^{\Theta(\cryptoParam)}$. This is used to track transaction commitments made by users. Transaction commitments are mapped from a unique piece of information which is also linked to a secret commitment mapping, and as such, a set of tokens. This is used to ensure that if a transaction commitment is made, the only way to use the set of tokens linked to the secret commitment is to reveal the unique committed transaction.
\end{itemize}

With this notation, wallets contain a set of tokens and a mapping of secret commitments to tokens within those wallets. At initialization all tokens in the wallet are mapped from the 0 secret commitment. To submit a transaction $tx$ using some subset of tokens $v$ in the wallet $w$, users must submit $(\commSerialNum, \commRandomness)$ satisfying 2 requirements:

\begin{enumerate}
    \item Users must provide a signature, as in a basic externally-owned account, such that $v \in C^w_\textit{scrt}(\commit(\commSerialNum,\commRandomness))$.
    \item If $C_{TX}(\commSerialNum)$ is non-zero, it must be that $C_{TX}(\commSerialNum)=\commit(tx)$. This is used to ensure that the user adheres to any commitment that they have made. 
\end{enumerate}

If either of these requirements are not met, the transaction is invalid. 
At the end of a transaction, the user must generate new mappings for all unmapped tokens in the wallets, with the default being the 0 secret commitment. If $C^w_\textit{scrt}(\commitment=\commit(\commSerialNum||\commRandomness))=v$ for some set of tokens $v$, $v$ is fixed (tokens cannot be added or removed from $v$) until $\commSerialNum$ is revealed. This is crucial in preventing a player from committing to a transaction which sends tokens without the player owning those tokens. If a player commits to bidding in an auction without any tokens in their wallet, but can add them before the tokens are needed, the player is able to effectively only reveal bids when bids are favourable.

\subsection{Relayers}\label{sec:Relayers}

A fundamental requirement for transaction submission in blockchains is the payment of some transaction fee to simultaneously incentivise block producers to include the transaction, and to prevent denial-of-service/spamming attacks. However, in account-/wallet-based models, this allows for the linking of player transactions, balances, and their associated transaction patterns. 

To counteract this, we utilise the concept of \textit{transaction relayers}, such as those used in $0$x\footnote{$0$x \url{https://0x.org/docs/guides/v3-specification}}, Open Gas Station Network\footnote{Open Gas Station Network \url{https://docs.opengsn.org/}}, and Biconomy\footnote{Biconomy \url{https://www.biconomy.io/}}. When a user wishes to submit a transaction anonymously to the blockchain, the user publishes a proof of membership in the set of registered users to the relayer mempool, as well as the desired transaction and a signature of knowledge cryptographically binding the membership proof to the transaction, preventing tampering. 

Generalizing the concept of a relayer, we assume that there exists a set of transactions $TX$, such that for any $\commit(tx)$ with $tx\in TX$, there exists a set of tokens described by $fee$, and a set of tokens described by $\textit{collateral}$, such that:

\begin{enumerate}
    \item $tx$ commits to send $fee$ to the relayer when $tx$ is revealed on the blockchain.
    \item $tx$ commits to burning $collateral$ if $tx$ is not revealed.
\end{enumerate}

For such a $tx$ with $fee$ and $\textit{collateral}$ big enough, a relayer is incentivized to include $\commit(tx)$ in the blockchain. Importantly, in this description, there is no requirement for the relayer to know $tx$, only $\commit(tx)$, $fee$ and $collateral$. Considering the cost for a user not revealing $tx$ is strictly greater than revealing due to $collateral$, relayers are incentivized to participate.

\section{PACCs}\label{sec;protocol}

In this section we describe PACCs and how they can be constructed using existing blockchain functionalities. We then outline some basic properties of PACCs. 

\subsection{Model}\label{sec:Model}

To reason about the properties of PACCs when applied to MEV, we introduce the following assumptions.

\begin{enumerate}
    \item A transaction submitted by a player for addition to the blockchain while observing blockchain height $\height$, is finalized in a block of height at most $\height+\revealTXDelay$, for some known $\revealTXDelay>0$.
    \item The public ZKP parameters (i.e. the keys $(pk, vk)$) are set-up in a trusted manner.\footnote{An example of such a set-up is a Perpetual Powers of Tau ceremony, as used in Zcash \url{https://zkproof.org/2021/06/30/setup-ceremonies/}}
    \item External market prices exist for all tokens, and follow Martingale processes.
    \item There exists a population of arbitrageurs able to frictionlessly trade at the external market price, who continuously monitor and interact with the blockchain.
    \item All players in the system are represented by a wallet located on a single blockchain-based distributed ledger, and a corresponding PKI.
\end{enumerate}

\subsection{PACCs Framework} \label{sec:conditions}

In the PACCs framework, there are two types of entity: players controlling wallets and relayers. 
Consider a player $\player_i$ wishing to add $\commit(tx)$ to the blockchain without anyone knowing that $\player_i$ generated $tx$, or exactly what $tx$ is. We emphasize exactly as we believe it is still important that $\player_i$ convinces a relayer that $tx$ is executable, and that $tx$ pays the relayer upon execution, as discussed in Section \ref{sec:Relayers}. To do this we introduce the concept of a \textit{break point} with respect to a transaction. A break point is such that for a transaction included for execution on a blockchain with a break point, the transaction executes every valid command up until the break point, regardless of what follows the break point. Thus, inserting a break-point can be considered as splitting up a transaction into two consecutive sub-transactions.
With this, we define \textit{base PACC transactions}.

\begin{definition}
    A transaction $tx$ is a \textit{base PACC transaction} if $\commit(tx)=\commit(tx^1$ $||tx^2)$ for $tx^1 $ equivalent to: \textit{pay $fee$, break-point}. The set of all base PACC transactions is denoted by $T_{base}$. Given the indistinguishability of base PACC transactions given only $\commit(tx)$ and a knowledge of the prefix $tx^1$, we consider $fee $ to be independent of the transaction contents $ \forall $ $ tx \in T_{base}$. 
\end{definition}

Recall that given a set of commitments $\commitments$ as specified in Section \ref{sec:ZK}, any valid ZKP from a player corresponding to $\commitment=\commit(\commSerialNum || \commRandomness)$ must reveal $\commSerialNum$. As such, consider a set of players $\player_1,...,\player_k$ owning wallets $w_1,...,w_k$ who create a single non-zero secret commitment mapping mapping for all of the tokens in their respective wallets. Let these mappings be of the form $\commitment_i=\commit(\commSerialNum_i || \commRandomness_i)$ for privately generated values $\commSerialNum_i$ and $ \commRandomness_i$ for each $\player_i$. We place the restriction on $\commSerialNum_i$ that there exists a $rootKey_i$, with $(rootKey_i, \commSerialNum_i)$ a valid private key, public key pair in the pre-defined PKI for $C_{TX}$. Typically, this means $\commSerialNum_i$ is derived from $rootKey_i$.

With this, we have enough to ensure relayers add transaction commitments on-chain within $\revealTXDelay$ blocks. Specifically, relayers will add transaction commitments to the global transaction commitment mapping, $C_{TX}$.
To demonstrate this, we outline a protocol which can be run between relayer and a wallet $w$ wishing to insert a transaction commitment $\commit(tx)$ to the blockchain.  

\subsection{PACCs Circuit} \label{sec:conditions}

\begin{figure}
  \centering
  \includegraphics[width=250px]{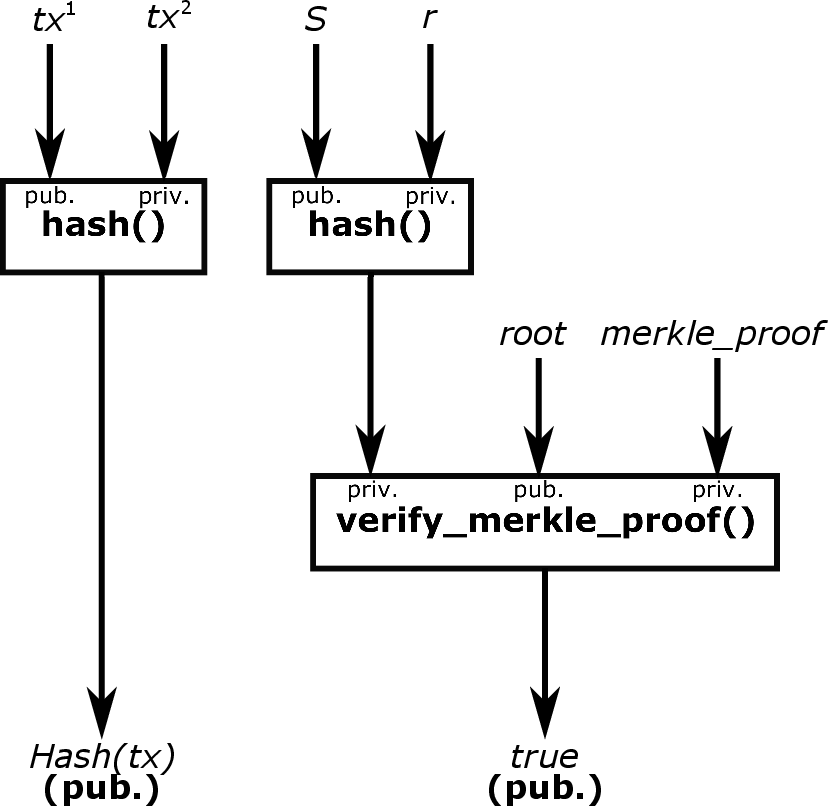}
  \caption{Arithmetic circuit for proving the user's availability to commit.}
  \label{fig:circuit}
\end{figure}

To participate in the protocol, retail users privately generate two bit strings, the \textit{serial number} $\commSerialNum$ and the \textit{randomness} $\commRandomness$, with $\commSerialNum, \ \commRandomness \in \{0,1\}^{\Theta(\cryptoParam)}$. Plus, they need to commit to two transactions $tx^1, tx^2$. Figure \ref{fig:circuit} depicts the arithmetic circuit that users have to compute $\pi \leftarrow$ \textit{Prove}$(pk, inputs)$ from, where $inputs = [tx^1, tx^2, S, r, root, merkle\_proof]$.

\subsection{PACCs Protocol}

Consider the set of all wallets $W=\{w_1,...,w_n\}$. Although PACCs allow for multiple secret commitments per wallet, WLOG, let the set of secret commitments $\commitments=\{\commitment_1,...,\commitment_n\}$ be such that $C^{w_i}_\textit{scrt}(\commitment_i)=bal(w_i),$ $\forall \ i \in [1,...,n]$. That is, there is one secret commitment per wallet, with the secret commitment mapping for each wallet mapping a single secret commitment to all of the tokens in that wallet. 

Let $W_b \subseteq W$ be the set of wallets with $bal(w_b) \geq fee+ collateral,$ $\forall \ w_b \in W_b $, and $\commitments_b$ be the secret commitments corresponding to these wallets. Specifically, $C^{w}_\textit{scrt}(\commitment_b)\geq fee+ collateral$, $\forall \commitment_b \in \commitments_b $. Therefore for a wallet $w_i$ with secret commitment $\commitment_i$,  $\commitment_i \in \commitments_b$ if and only if $w_i \in W_b$. This implies such a wallet $w \in W_b$ can produce a valid ZKP for the PACCs circuit, where the inputs' Merkle proof is computed as $merkle\_proof = \textit{SetMembership}(\textit{Hash}(\commRandomness, \commSerialNum), \commitments_b)$.

Furthermore, consider a wallet $w_i \in W_b$ wishing to insert a commitment to a transaction $tx_i \in T_{base}$ into the global transaction commitment mapping. As $tx_i\in T_{base}$, this implies $tx_i=tx^1||tx^2$ with $tx^1$ equivalent to: pay $fee$, break point. This means $w_i$ can produce a valid ZKP for the PACCs circuit, where the preimage inputs are $(tx^1, tx^2)$. As this ZKP reveals $tx^1$ and proves knowledge of the whole preimage of $\commit(tx_i)$, it must be that $tx^1$ is a prefix of $tx_i$, without revealing anything else about $tx_i$. 

Therefore, let $w_i$ send this ZKP, and a signature of this message using $rootKey_i$. This signature can be verified using $\commSerialNum_i$, which is revealed by the ZKP. This further ensures the player generating $\commit(tx_i)$ must also have generated $\commSerialNum_i$, under standard PKI assumptions. 

From the Merkle proof verification in the ZKP, the relayer knows that $bal(w_i) \geq fee+ collateral$, as such a proof is only possible if  $w_i \in W_b$. Accompanied with the transaction hash of the ZKP, the relayer then knows:

\begin{enumerate}
    \item $w$ commits to send $fee$ to the relayer when $tx_i$ is executed.
    \item $w$ commits to burning $collateral$ if $tx_i$ is not executed. This is because the relayer knows that at least $collateral$ exists in the wallet. Furthermore, as $\commSerialNum_i$ is mapped to $\commit(tx_i)$ in the global transaction commitment mapping $C_{TX}$, by definition of $C_{TX}$, only revealing $tx_i$ allows $w_i$ to use the tokens mapped from $\commitment_i= \commit(\commSerialNum_i || \commRandomness_i)$ in $C^{w_i}_\textit{scrt}()$.  As such, if $tx_i$ is never revealed, at least $collateral$ is burned in $w_i$.
\end{enumerate}

\subsection{Full Protocol}

\begin{figure*}
  \centering
  \includegraphics[width=\textwidth]{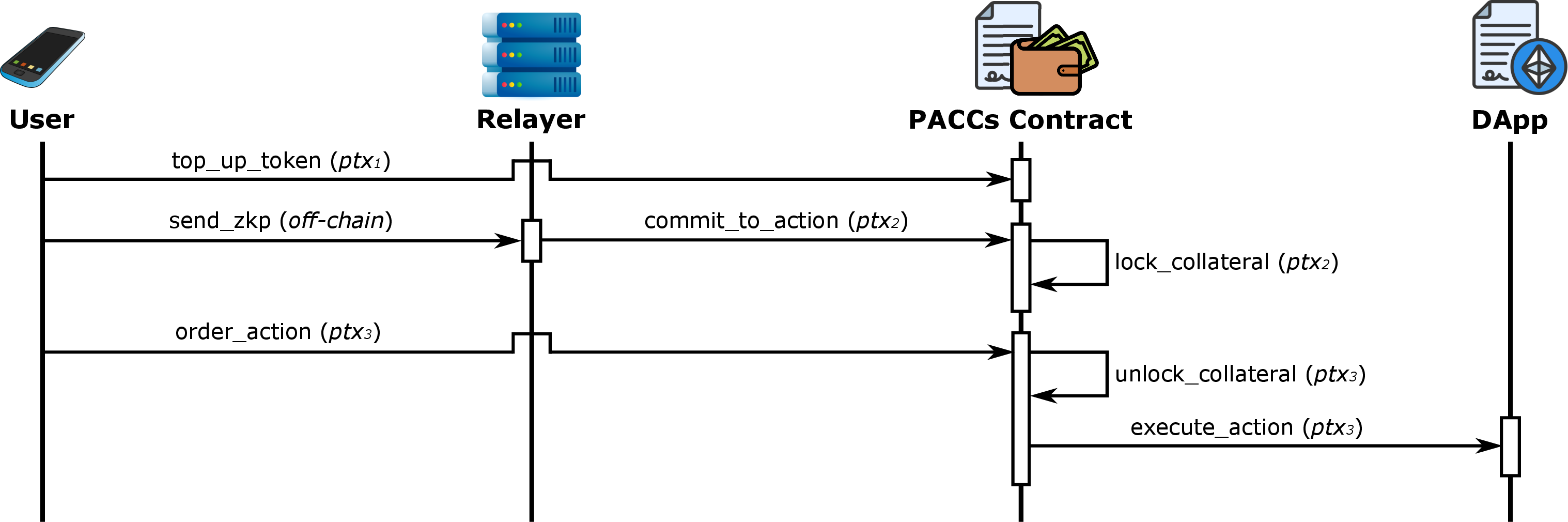}
  \caption{Overview of the protocol steps performed between the user, the PACCs contract, and the DApp, in order to peform an action on the DApp.}
  \label{fig:protocol}
\end{figure*}

In this section, we outline the full protocol involving the proposed PACCs solution. Let us have a user (i.e. an EOA) willing to use a DApp (e.g. a DEX), the DApp, and our PACCs contract. In order to perform the full protocol to activate an action on the DApp, a total of three transactions are required: ($ptx_1, ptx_2, ptx_3$). How these are issued and the steps they perform in order to perform an action, are the following (and are depicted in Figure \ref{fig:protocol}):

\begin{itemize}
    \item \textbf{(user)} \verb!top_up_token! ($ptx_1$): send an amount of a given token to the PACCs contract, along with a commitment $\textit{Hash}(\commRandomness, \commSerialNum)$. Once the token is received, if $bal(w_i) \geq fee + collateral$, the contract updates the user account with the new amount and adds the provided commitment to the contract state, into a Merkle tree of commitments. At this point, the commitment publicly belongs to "someone" having enough funds to pay for the service fee and the collateral. Plus, the tokens can only be spent if the opening to the commitment is revealed.
    
    \item \textbf{(user)} \verb!send_zkp! (\textit{off-chain}): when a DApp action wants to be performed, the user first needs to commit to such action. As such, the user sends a PACCs ZKP to the relayer properly signed using a $rootKey$.

    \item \textbf{(relayer)} \verb!commit_to_action! ($ptx_2$): if the ZKP verifies, and the signature can be verified using $S$, the relayer forwards the commitment $\commit(tx_i)$ and the value $S$ to the PACCs contract.
    
    \item \textbf{(PACCs contract)} \verb!lock_collateral! ($ptx_2$): upon receiving the commitment, the collateral gets locked in the contract. In particular, it places a restriction where the opening to the commitment $\commit(tx_i)$ needs to be revealed by who committed to $\textit{Hash}(\commRandomness, \commSerialNum)$ using $S$.
    
    \item \textbf{(user)} \verb!order_action! ($ptx_3$): after some time, the user orders the action, by issuing the promised transaction, thus revealing $tx^2$. Plus, the previous $r$ opening is revealed, and also a fresh new commitment $\textit{Hash}(\commRandomness, \commSerialNum)$ is provided.

    \item \textbf{(PACCs contract)} \verb!unlock_collateral! ($ptx_3$): if everything worked with no aborts, the collateral gets unlocked by removing the restriction.
    
    \item \textbf{(PACCs contract)} \verb!execute_action! ($ptx_3$): if the received order was indeed committed previously (i.e. the transaction itself is correct), the action is executed by calling the DApp. Plus, the commitment stored in the PACCs contract state gets replaced by the new one (or the existing one gets deleted if not enough funds remain to perform a new exchange). In this step, the fee is paid to the relayer as well, for the service they provided.

\end{itemize}

\section{Implementation and Benchmarks}\label{sec:im}

The PACCs contract is the core element of our protocol. We implemented it using Solidity\footnote{https://soliditylang.org/}, and deployed it on an Ethereum devnet using the Foundry\footnote{https://github.com/foundry-rs/foundry} toolkit. The contract implements the basic features of an standard SCW, with the added functionalities that relate to our protocol. Ideally, such contract is deployed by a party trusted by the DApps interested in using it, or the same DApp owner. Such a contract handles the accounts of all users, and interacts directly with the desired DApps. Likewise, we implemented a contract simulating a basic DApp, a DEX where an ERC20 token is provided to users in exchange for Ether. We measured the gas consumption for the three transactions involved in our protocol, achieving similar results compared to most smart contract functionalities (e.g. simply transferring Ether to an address is 21000 gas): 

\begin{itemize}
    \item $ptx_1$: 34221 gas
    \item $ptx_2$: 27567 gas
    \item $ptx_3$: 53415 gas
\end{itemize}

Nevertheless, it must be taken into account that our PoC has been implemented directly on Ethereum, considering the worst-case scenario. In other words, optimizations like batching (e.g. including many transactions into a single one using technologies like zk-Rollups \cite{9862815}) or other scaling solutions like Polygon\footnote{https://polygon.technology/} make our protocol completely feasible in terms of usage cost.

On the other hand, the user needs to generate ZKPs, and the relayer needs to verify them, being the proof generation the most expensive operation in our protocol, in terms of computing resources. In such regard, we use \textit{dusk-plonk}\footnote{https://github.com/dusk-network/plonk}, a Rust crate to write circuits for the Plonk proof system \cite{cryptoeprint:2019/953}, that allows also to generate and verify them. Plonk is commonly used in production, and serves our purposes.

The circuit we wrote uses the Poseidon hash \cite{cryptoeprint:2019/458} function as a commitment scheme, an efficient hash ideal for in-circuit operations. We wrote the PACCs circuit using an anonymity set of $4^{14}$ elements (i.e., two 4-arity Merkle trees of depth 14), which leads to a circuit of \textbf{15889 constraints}. We benchmarked it using an AMD Ryzen 7 5800X CPU, taking only \textbf{1.610 seconds} to generate the proof, and just \textbf{0.006 seconds} to verify it. This last metric is important for what concerns the relayer, because the amount of workload introduced to it by the ZKP is almost negligible. And for what concerns the prover, we can see how the achieved result is excellent as well, as performing the protocol in a matter of seconds is totally feasible for the user.

\section{Properties of PACCs}\label{sec:properties}

Towards applying PACCs to MEV, we detail some of properties that PACCs possess. The first result is that, within the competition assumptions of Section \ref{sec:Model}, the expected collateral required to submit a PACC diminishes to 0.

\begin{lemma}\label{lem:diminishingCol}
    For a base PACC transaction $tx$ from a wallet $w$ with $bal(w)>fee$, the Nash Equilibrium for $collateral=0$.
\end{lemma}

\begin{proof} 
    For any wallet $w$ posting a commitment to such a transaction $tx$ with $bal(w)>fee$, the payoff for revealing $tx$ is at least $collateral$, compared to $0$ for not revealing. Therefore, any collateral greater than 0 is sufficient to incentivize revelation. Given a cost to $w$ for locking collateral, the optimal collateral for $w$ is 0. Therefore, $w$ posts the minimum possible collateral, which approaches 0. 
\end{proof}

\subsection{{PACCs} and MEV}\label{sec:sims}

With the core PACCs protocol described in Section \ref{sec;protocol}, we are able to shift almost all known MEV to censorship. Given competing block producers and transaction fees, censorship can be made arbitrarily expensive and unlikely, as the cost to bribe block producers to censor is at most $O(\revealTXDelay)$ \cite{proposercensorshipResnick}. This section applies PACCs to the main sources of MEV starting with how PACCs can be leveraged to run a sealed-bid auction. We then demonstrate the potential of PACCs in tackling MEV, focusing on how PACCs enable users and/or protocols to buy or sell tokens at relevant external market prices, excluding transaction fees. This is analogous to MEV prevention in most documented sources of MEV \cite{FlashBoys2.0,QuantifyingExtractableValueGervais,DeFiLiquidationsGervais}.

\begin{lemma}\label{lem:PACCAuctions}
    Sealed-bid auctions can be run among all wallets $W$ such that $bal(w)>fee, \ \forall \ w \in W$. 
\end{lemma}

\begin{proof} 
    Consider such a wallet $w \in W$, and an on-chain auction $A$. Let $A$ accept bid commitments for $\revealTXDelay$ blocks, followed by $\revealTXDelay$ blocks in which bids committed to $A$ can be revealed.
    Consider a base PACC transaction $tx$, with $tx_2 \equiv \textit{bid X in auction }A$ for some set of tokens $X$. We know for any $w \in W$ committing to $tx$ and proving membership in $W$, a relayer adds $\commit(tx)$ to the blockchain within $\revealTXDelay$ blocks. Therefore any $w \in W$ submitting this bid commitment and proofs to a relayer is included in $A$. Put differently, after the initial $\revealTXDelay$ blocks, any player who wanted to commit a bid for $A$ has been included in the blockchain. 
    
    After the initial $\revealTXDelay$ blocks, all players reveal their bids. As bids can be revealed for up to $\revealTXDelay$ blocks, this is sufficiently long for all players revealing bids to be added to the blockchain. As the committed transactions are base PACC transactions, all players committing to a bid are incentivized to reveal. Importantly, all bids were committed before any information regarding other bidders was revealed, with revealed bids matching committed bids due to the requirement of the global transaction commitment mapping for revealed transactions from $w$ to match any transaction committed by $w$.
    
    Thus, at the end of $\revealTXDelay$ blocks, all bids are revealed on chain. Any player in the blockchain can then settle the auction, deducing the auction winner and settlement price from the revealed bids. This is sufficient to run a sealed-bid auction. 
\end{proof}

\subsubsection{Liquidations}

Lemma \ref{lem:PACCAuctions} has immediate consequences for MEV in liquidations. Such MEV is highlighted empirically in \cite{DeFiLiquidationsGervais}. The source of this MEV is the ability for players to trigger liquidation of collateral and buy the collateral at a discount. To address this, we can replace the \textit{Auction Liquidations} as labelled in \cite{DeFiLiquidationsGervais} with sealed-bid auctions based on PACCs.
Using a result from auction theory on sealed-bid auctions among competing players with the same view of external market prices\cite{krishna2009auction}, we know that seller revenue of such an auction is at least the value of the tokens being sold. As PACCs allow player with tokens to participate in any auction, PACCs also allow for such a sealed-bid auction among competing player. Given the external market prices of these tokens follow Martingales with which arbitrageurs can interact with frictionlessly (Section \ref{sec:Model}), the expected revenue of a liquidation auction when the auction starts is at least that of the external market value of the collateral when the auction starts, as required.

Finally, a subtle yet important property of PACCs is that a commitment to a valid send transaction requires the tokens to be present in the sending wallet when the commitment is made. Specifically, for a wallet $w$ with  $C^{w}_\textit{scrt}(\commit(\commSerialNum,\commRandomness))=v$ for some set of tokens $v$ and any transaction $tx$ from $w$ sending tokens to another smart contract or wallet, if $C_{TX}(\commSerialNum)=\commit(tx)$ before $tx$ is revealed, for $tx$ to be valid, it must be that there were enough tokens in $v$ when the secret commitment map was created. This is by definition of $C^{w}_\textit{scrt}()$, with only tokens mapped from $\commSerialNum$ being usable in $tx$.

\subsubsection{Decentralized Exchange}

Decentralized exchange is the primary source of MEV in current blockchains ($>99\%$ of the MEV identified as extracted by \cite{FlashbotsExploreWebsite}, as seen in the chart labelled ``Extracted MEV Split by Type" ). From the same principles as Lemma \ref{lem:PACCAuctions}, PACCs can implement a significantly more powerful auction with respect to decentralized exchange, a frequent batch auction (FBA) \cite{FrequentBatchAuctionsBudish}. FBAs are auctions designed to connect buyers and sellers of a particular token/set of tokens, and are proven to settle at external market prices under competition among MMs. A decentralized variant of an FBA is implemented in the FairTraDEX protocol \cite{FairTraDEXMcMenamin}. FairTraDEX demonstrated that implementing FBAs was possible in a decentralized setting without the trusted auctioneer required in \cite{FrequentBatchAuctionsBudish}. 

\begin{corollary}\label{cor:PACCFBAs}
  Frequent batch auctions can be run among all wallets $W$ such that $bal(w)>fee, \ \forall \ w \in W$. 
\end{corollary}

\begin{proof} 
   From Lemma \ref{lem:PACCAuctions}, we know we can implement any auction involving sealed-bids in any set of tokens. For any pair of token sets $(X,Y)$, consider an auction $A$ where players submit bids as base PACC transactions, of the form $tx_1||tx_2$, with $tx_2$ equivalent to \textit{buy $v_x>0$ of $X$ for $v_y>0$ of $Y$} or \textit{buy $v_y>0$ of $Y$ for $v_x>0$ of $X$}. If the auction is run over $2\revealTXDelay$ blocks, this is enough for players to commit sealed bids to the blockchain, and have them revealed. From \cite{FairTraDEXMcMenamin}, we can implement an on-chain clearing-price verifier to ensure orders of this format get executed at a unique clearing price which maximizes traded volume, and as such, replicates a frequent batch auction. 
\end{proof}

FairTraDEX requires bids to be collateralized, indistinguishable, and committed to the blockchain before being revealed and settled at a unique clearing price. PACCs allow for all of these requirements, while providing significant improvements on the FairTraDEX protocol. Players using PACCs can participate in FBAs in any token/token-set pair at any time (in addition to optionally performing normal transactions without additional on- or off-boarding), compared to players in FairTraDEX being restricted to FBAs in one trading pair. With PACCs, players can immediately join any ongoing FBA. With the anonymity set being the set of wallets, there is never a need to diffuse or on-board other players for an auction. Furthermore, PACCs remove the need to perform Merkle-Tree inserts and verifications on-chain, making PACCs significantly cheaper and more scalable.

\subsubsection{PACC RFQ protocol} 

Despite the promises of FBAs, adoption is limited. With PACCs, there is potential for other DEX alternatives with similar price guarantees, which can be seen as more aligned with the desired experience of retail users. 
In this regard, we consider first a request-for-quote (RFQ) style DEX protocol based on PACCs, which resembles a fee-escalator as proposed in an Ethereum Improvement Proposal\cite{FeeEscalatorEIP}. 

For a swap between two sets of tokens $X$ and $Y$ with external market price $\MIFP$, specifically $\MIFP X=Y$ , consider a PACC from a wallet $w$ proving membership in a set of wallets with \textit{at least $X$ tokens or $Y$ tokens}. Consider $f_H: \mathbb{N}_{\geq H} \rightarrow \mathbb{R}_{\geq 0}$, an increasing function with $f_H(h)$ undefined for $h<H$. The function $f_H(h)$ defines a commission to be paid by the user to the relayer including their transaction in the chain at block height $h$. 

The accompanying transaction commitment is proved to be a member of the set $\{com_1, com_2\}$, with:
\begin{itemize}
    \item $com_1=\commit(tx_1=tx_a||tx_b)$ and $tx_a \equiv$ \textit{sell $X$ for $ Y$, pay $f_H(h)$ to relayer, break point},
    \item $com_2=\commit(tx_2=tx_c||tx_d)$ and $tx_c \equiv$ \textit{sell $ Y$ for $X$, pay $f_H(h)$ to relayer, break point}.
\end{itemize}

A relayer adding the commitment to the blockchain must collateralize the order with at least $X$ and $Y$ to be valid, with the relayer trading with $w$ when the committed transaction is revealed. After $\revealTXDelay$ blocks from when $w$'s transaction is committed to the blockchain, the relayer can reclaim their own collateral. After this point, $w$'s tokens are burnt if $w$ reveals the committed transaction, or locked indefinitely if not.

\begin{lemma}\label{sec:RFQ}
    Given no transaction fees, the PACC RFQ protocol has Nash Equilibrium involving MMs committing user orders to the blockchain with $E(f_H)=0$.
\end{lemma}

\begin{proof}
   Firstly, given the order is generated when the external market price is $\MIFP$, for any block generated after this point (after block $H$), the expected external market price is $\MIFP$ due to the Martingale assumption. As all MMs observe $\MIFP$, the expected revenue for a MM responding to a PACC RFQ is at least 0 and strictly increasing in block height, making this a Dutch Auction in the commission specified by $f_H()$ between MMs. By the revenue equivalence principle \cite{krishna2009auction}, the expected revenue for the seller is the same as if this were an sealed-bid auction to receive $f_H()$. For any positive value of $f_H$, MMs will bid for this opportunity. As the user chooses $f_H()$, the commission to pay, but does not receive the revenue of the auction, for any positive value for $f_H(h)$, $h \geq H$, the user is strictly incentivized to reduce $f_H(h)$. As such, the Nash Equilibrium of $E(f_H)$ is $E(f_H)=0$. 
\end{proof}

\begin{corollary}\label{prop:RFQ}
    Given no transaction fees, the PACC RFQ protocol has Nash Equilibrium in which users trade at the external market price. 
\end{corollary}

This protocol has many of the expected price benefits of an FBA based on PACCs, with users expecting to trade at the external market price excluding fees in both cases. PACC RFQs come with the added benefit for users that users are only required to reveal when the order has been executed. Depending on the MM preferences or requirements, PACC RFQs can also be used to enforce Know-Your-Customer and/or anti-money laundering checks by MMs before responding, with MMs able to require arbitrary membership proof rules. For example, if European MMs choose to only respond to wallets who have received a verification token from European regulators, PACCs allow users to preserve anonymity within this set, while maintaining the privacy required to trade at the external market price in expectancy. 

As mentioned, this resembles a fee escalator \cite{FeeEscalatorEIP}, with the strict benefit of not revealing trade price, direction, or identity to searchers before the trade has been committed to. All of these pieces of information together can be enough to move the external market price before the order gets interacted with. If searchers see a protocol creator selling protocol tokens, this could have significant negative sentiment effects on the price of the token. Therefore, the searchers might pre-emptively move the price when such an order enters the mempool. In contrast, if the same player committed to either buying or selling tokens using a PACC, with a membership proof in a group equally likely to buy or sell, the user expects to trade around the external market price.

\subsubsection{PACCs and AMMs} 

After the highlighting of the phenomenon of loss-versus-rebalancing \cite{LVRRoughgarden}, we expect block producers to arbitrage an AMM price to the external market price. A recent AMM proposal using ZK commitments \cite{V0LVERMcMenamin} provides an interesting use case for PACCs. Consider then an AMM, such as that introduced in \cite{V0LVERMcMenamin} which accepts PACCs. Given the delay between commitment and revelation, it seems necessary that the AMM is required to lock-up reserves to trade with user orders. For this to be viable from an AMMs standpoint (opportunity costs are added), the AMM should require some fee to be deposited by a relayer allocating funds for a user order (which can be incorporated by the relayer into the fee/collateral required from a PACC). If all orders outside of the block producer arbitrage are executed at the same price, a user should expect their order, at worst, to be executed at the external market price at time of commitment, minus the fees and impact for interacting with the AMM. As the arbitrage from the producer is unaffected, all PACC orders paying inclusion fees should be added to the chain. 

Current AMM users only expect their orders to trade after a searcher has moved the price of the AMM, meaning interacting with the AMM at the external market price is the best-case scenario for users currently. Given the extent of user-level MEV \cite{FlashbotsExploreWebsite,QuantifyingExtractableValueGervais}, this best-case scenario has not been considered attainable. As such, PACCs have the potential to drastically reduce MEV in AMMs. Following recent work on LVR-proofing AMMs \cite{McAMMs,DiamondMcMenamin}, PACCs may be pivotal in making AMMs LVR- and MEV-proof.

\section{Conclusion}\label{sec:conclusion}

We outline PACCs, a protocol allowing anyone with sufficient capital to anonymously and privately commit collateralized transactions for any protocol to a blockchain. This is compared to earlier solutions based on anonymity sets \cite{EagleBaum,FairTraDEXMcMenamin,P2DEXBaum} that force players trying to achieve the similar levels of anonymization and privacy for collateralized commitments to join anonymity sets in advance of the opportunity, typically before it exists. This necessity to lock up capital at some, potentially significant, opportunity cost limits the applicability of such solutions. 

The trade-off with PACCs is the dependency on relayers to post collateral on behalf of unknown players, with only game-theoretic guarantees of repayment. We see this as an acceptable trade-off, introducing some unpredictability with respect to when and if rewards are paid out. Importantly, collateral and fees from wallets can be enforced to reflect this unpredictability, and given competition among relayers for these fees, the equilibrium for these fees should approach the gas costs paid by relayers for including the transaction as indicated in Lemma \ref{lem:diminishingCol}.

Although we propose PACCs for use in commit-reveal protocols, mainly due to their provable decentralization, and ability to be implemented immediately on any smart-contract enabled blockchain, there are several alternatives that are worth mentioning. In our description of PACCs, the relayer includes transaction commitments to the blockchain. These commitments can be replaced by a threshold encryption of the transaction \cite{Penumbra,HelixAsayag}, or using a delay encryption scheme \cite{DelayEncryptionBurdges}. In these encryption schemes, it may be possible to reduce collateral requirements, although practical and decentralized variations of these schemes have yet to be proposed. Improvements in these areas will greatly improve the usability and capabilities of PACCs. Importantly, all schemes, including the current description of PACCs, have a distinct committal of information to the chain, followed by a revelation of information. As such, commit-reveal accurately describes the process taking place, regardless of the specific revelation scheme being used.

\bibliographystyle{plain}
\bibliography{references}


\end{document}